\documentstyle[11pt,springer1]{article}

\begin{document}
\input{psfig.sty}

\renewcommand{\textfraction}{0.01}
\renewcommand{\topfraction}{0.99}
\renewcommand{\bottomfraction}{0.99}
\renewcommand{\floatpagefraction}{0.99}
\setcounter{topnumber}{5}
\setcounter{bottomnumber}{5}


\newcommand{\nbcite}[1]{\citeauthor{#1},\ \citeyear{#1}}


\newenvironment%
     {itemizes}%
     {\vspace{-2mm}\begin{itemize}\addtolength{\itemsep}{-2mm}}
     {\end{itemize}}

\newenvironment%
     {enumerates}%
     {\vspace{-2mm}\begin{enumerate}\addtolength{\itemsep}{-2mm}}
     {\end{enumerate}}

\newtheorem%
     {theorem}{Theorem}[section]
\newtheorem%
     {corollary}[theorem]{Corollary}
\newtheorem%
     {proposition}[theorem]{Proposition} 
\newtheorem%
     {lemma}[theorem]{Lemma} 

\newtheorem%
     {exampleAux}[theorem]{Example}

\newenvironment%
     {example}{\begin{exampleAux}\rm}{\end{exampleAux}}

\newtheorem%
     {examplesAux}[theorem]{Examples} 
\newenvironment%
     {examples}{\begin{examplesAux}\rm}{\end{examplesAux}}

\newtheorem{definition}[theorem]{Definition}

\newtheorem%
     {constructionAux}[theorem]{Construction} 
\newenvironment%
     {construction}{\begin{constructionAux}\rm}{\end{constructionAux}}

\def\Proof{{\sl Proof.\ }}

\newenvironment%
     {proof}{\noindent\Proof }{\qed}

\def\qed{\hfill{\boxit{}}
  \ifdim\lastskip<\medskipamount \removelastskip\penalty55\medskip\fi}
\long\def\boxit#1{\vbox{\hrule\hbox{\vrule\kern3pt
                  \vbox{\kern3pt#1\kern3pt}\kern3pt\vrule}\hrule}}

\newenvironment%
     {genass}%
        {\medbreak\noindent{\bf General Assumption.\enspace}\it}%
        {\ifdim\lastskip<\medskipamount \removelastskip\penalty55\medskip\fi}


\newcommand {\boxfigure}[1]%
   {\framebox[\textwidth]{%
    \parbox {0.99\textwidth}
                {{#1}\vspace {0cm}\hfill}}}

\newcommand {\boxfigureone}[1]%
   {\framebox[\textwidth]{%
    \parbox {0.90\textwidth}
                {{#1}\vspace {0cm}\hfill}}}

\newcommand{\gefig}[3]
        {\begin{figure}[htb] 
        \begin{center}%
        \mbox{}
        {\psfig{figure=#1,width=0.60\textwidth}}
        \mbox{}
        \end{center}
        \caption{{#2}\label{#3}} 
        \end{figure}}

\newcommand{\sizefig}[4]
        {\begin{figure}[htb] 
        \begin{center}%
        \mbox{}
        {\psfig{figure=#1,width=#2}}
        \mbox{}
        \end{center}
        \caption{{#3}\label{#4}} 
        \end{figure}}

\newcommand{\sizefigstar}[4]
        {\begin{figure*}[htb] 
        \begin{center}%
        \mbox{}
        {\psfig{figure=#1,width=#2}}
        \mbox{}
        \end{center}
        \caption{{#3}\label{#4}} 
        \end{figure*}}

\newcommand{\hangif}[1]{\raisebox{-1ex}{\hspace{2em}
                                        \makebox[1.0em][l]{if}
                                        \parbox[t]{320pt}{#1}}}



\def\A{{\cal A}} \def\B{{\cal B}} \def\C{{\cal C}} \def\D{{\cal D}}
\def\E{{\cal E}} \def\F{{\cal F}} \def\G{{\cal G}} \def\H{{\cal H}}
\def\I{{\cal I}} \def\J{{\cal J}} \def\K{{\cal K}} \def\L{{\cal L}}
\def\M{{\cal M}} \def\N{{\cal N}} \def\O{{\cal O}} \def\P{{\cal P}}
\def\Q{{\cal Q}} \def\R{{\cal R}} \def\S{{\cal S}} \def\T{{\cal T}}
\def\U{{\cal U}} \def\V{{\cal V}} \def\W{{\cal W}} \def\X{{\cal X}}
\def\Y{{\cal Y}} \def\Z{{\cal Z}}

\newcommand{\dd}[2]{#1_1,\ldots,#1_{#2}}      

\newcommand{\set}[1]{\{#1\}}
\newcommand{\eset}{\emptyset}
\newcommand\bigset[1]{ \Bigl\{ #1 \Bigr\} }   
\newcommand\bigmid{\ \Big|\ }

\newcommand{\incl}{\subseteq}		
\newcommand{\incls}{\supseteq}		

\newcommand{\col}{\colon}

\newcommand{\NP}{{\rm NP}}		
\newcommand{\GI}{{\rm GI}}              
\newcommand{\PIPEETWO}{\Pi^{{\rm P}}_2}	
\newcommand{\SIGPEETWO}{\Sigma^{{\rm P}}_2}	
\newcommand{\PSPACE}{{\rm PSPACE}}	
\newcommand{\PTIME}{{\rm P}}		
\newcommand{\EXPTIME}{{\rm EXPTIME}}	

\newcommand{\angles}[1]{\langle#1\rangle}	


\newcommand{\OnlyIf}{\lq\lq$\Rightarrow$\rq\rq\ \ }   
\newcommand{\If}{\lq\lq$\Leftarrow$\rq\rq\ \ }        

\newcommand{\quotes}[1]{\lq\lq#1\rq\rq}     	
\newcommand{\wrt}{w.r.t.}			
\newcommand{\WLOG}{w.l.o.g.}			
\newcommand{\ie}{i.e.}		        	
\newcommand{\eg}{e.g.}		        	
\renewcommand{\hom}{homo\-mor\-phism}

\newcommand{\eat}[1]{}

\newcommand{\comment}[1]{\noindent{\sl COMMENT:} {\sl #1}}

\newcommand{\card}[1]{|#1|}
\newcommand{\AM}{{\sl AM}}
\newcommand{\WM}{{\sl WM}}
\newcommand{\vquant}{{\sl vquant}}
\newcommand{\ElementDef}[2]{<\!\!\!\!\!~\mbox{!ELEMENT ${#1}$   ${#2}$}\!\!>}


\newcommand{\PR}{{\sl Pr}}                      

\newcommand{\EXISTS}{{\sl One Is}}
\newcommand{\NOTEXISTS}{{\sl None Is}}
\newcommand{\FORALL}{{\sl All Are}}
\newcommand{\NOTFORALL}{{\sl Not All Are}}

\newcommand{\Count}{{\sl count}}
\newcommand{\Sum}{{\sl sum}}
\newcommand{\Min}{{\sl min}}
\newcommand{\Max}{{\sl max}}
\newcommand{\Avg}{{\sl avg}}
\newcommand{\None}{{\sl none}}
\newcommand{\MinCon}{{\sl min}${}^c$}
\newcommand{\MaxCon}{{\sl max}${}^c$}

\newcommand{\Set}{{\sl def}}

\newcommand{\Dollar}{\$}
\newcommand{\var}[1]{\Dollar #1}

\newcommand{\terminal}{\phi}
\newcommand{\rt}{{``root''}}
\newcommand{\assign}[1]{set(#1)}


\newcommand{\cond}{{\sl cond}}
\newcommand{\quant}{{\sl quant}}
\newcommand{\agg}{{\sl agg}}
\newcommand{\vardef}{{\sl var\_def}}

\newcommand{\extension}[1] {\overline{#1}} 

\newcommand{\aomap}{{\sl and/or}}

\newcommand{\ComputeQuery}{{\tt Compute\_Query}}
\newcommand{\BUComputeQuery}{{\tt Bottom\_Up\_Compute\_Query}}

\newcommand{\GraphOf}[1]{(N_{#1},E_{#1},r_{#1})}

\newcommand{\SatMat}[2]{{\M}^{#2}_{#1}}
\newcommand{\Wit}[3]{{\W}^{#3}_{#2}({#1})}

\newcommand{\negation}{{\sl negation}}

\newcommand{\val}[1]{{\sl values}({#1})}
\newcommand{\EquiXagg}{{\rm EquiX^{\rm agg}}}
\newcommand{\EquiXreg}{{\rm EquiX^{\rm reg}}}

\newcommand{\qao}{o}
\newcommand{\qef}{q}
\newcommand{\query}{(T,l,c,\qao,\qef, O)}
\newcommand{\aggquery}{(T,l,c,\qao,\qef, a, a_c, O)}
\newcommand{\queryOf}[1]{(T_{#1},l_{#1},c,\qao,\qef, O)}
\newcommand{\matches}{{\sl matches\/}}


\title{EquiX---A Search and Query Language for XML}
\author{Sara Cohen\thanks{Institute for Computer Science, The Hebrew University, Jerusalem 91904, Israel.} %
\and    Yaron Kanza${}^{*}$
\and    Yakov Kogan${}^{*}$
\and    Werner Nutt\thanks{German Research Center for Artificial Intelligence Gm
bH, Stuhlsatzenhausweg 3, 66123 Saarbr\"ucken, Germany}
\and    Yehoshua Sagiv${}^{*}$
\and    Alexander Serebrenik\thanks{Computer Science Dept.,
        K. U. Leuven, Heverlee, Belgium }}
\date{}
\maketitle






\begin{abstract}

EquiX is a search language for XML that combines the power
of querying with the simplicity of searching. Requirements for such languages
are discussed and it is shown that EquiX meets the necessary criteria.
Both a graphical abstract syntax and a formal concrete syntax 
are presented for EquiX queries.
In addition, the semantics is defined and an evaluation algorithm is presented.
The evaluation algorithm is polynomial under combined complexity. 

EquiX combines pattern matching, quantification and logical expressions
to query both the data and meta-data of XML documents.
The result of a query in EquiX is a set of XML documents. 
A DTD describing the result documents
is derived automatically from the query.

\eat{
This is particulary important when quering large repositories.

We present EquiX---a powerful, yet easy to use query language for XML. 
The main goal in designing EquiX is to strike the
right balance between expressive power and simplicity.

EquiX has a form-based GUI that is constructed automatically from the DTD of 
XML documents. Query forms are built from well known HTML primitives. 
The result of a query in EquiX is a collection of XML pages, and it is
automatically generated from the query without explicit specification 
of the format of the result. The DTD describing the result documents
is derived automatically from the query form.
We provide an algorithm for query evaluation that runs in polynomial time,
even when the query is considered as part of the input (i.e., combined complexity).
 
Knowledge of XML syntax is not required in order to use EquiX. Yet, EquiX
is able to express rather complicated queries. For example, it can 
express quantification, negation and aggregation.  
}

\end{abstract}

\section{Introduction} \label{Section-Introduction}

The widespread use of the World-Wide Web has given rise to a plethora of 
simple query processors, commonly called search engines. 
Search engines query a database of semi-structured data, namely HTML pages.
Currently, search engines cannot be used to query the meta-data content in such
pages. Only the data can be queried. For example, one can use a search engine
to find pages containing the word ``villain''. However, it is difficult to 
obtain only pages in which villain appears in the context of a character in 
a Wild West movie. More and more XML pages are finding their way onto the Web.
Thus, it is becoming increasingly
important to be able to query both the data and the meta-data content of 
the pages on the Web. We propose a language for querying (or searching) the
Web that fills this void.

Search engines can be viewed as simple query processors. The query language 
of most search engines is rather restricted. Both traditional database
query languages, such as SQL, and newly proposed languages, such as
XQL~\cite{XQL-Microsoft}, 
XML-QL~\cite{XML-QL} and Xmas~\cite{XMAS:W3C,XMAS:DTD:Inference},
are much richer than the query language of most search engines.
However, the limited expressiveness of search engines
appears to be an advantage in the context of the Web.
Many Internet users are not familiar with database concepts and find it hard
to formulate SQL queries. 
In comparison, when it comes to using search engines,
experience has proven that even novice Internet users can easily ask queries 
using a search engine. It is likely that this is true because of the inherent 
simplicity of the search-engine query languages.

Consequently, an apparent disadvantage of search-engine languages is really 
an advantage when it comes to querying the Web. Thus, it is 
imperative to first understand the requirements of a query language for the 
Web, before attempting to design such a language. We believe that the 
Web gives rise to a new concept in query languages, namely {\em search 
languages\/}. We will present design criteria for search languages.

As its name implies, a search language is a language that can be used to 
search for data. We differentiate between the terms {\em search\/} and 
{\em query\/}. Roughly speaking, a search is an imprecise process in which
the user guesses the content of the document that she requires. Querying
is a precise process in which the user specifies exactly the information she
is seeking. In this paper we define a language that has both searching and 
querying capabilities. We call a language that allows both searching and 
querying a search language.

We call a query written in a search language a {\em search 
query\/} and the query result a {\em search result\/}. Similarly, 
we call a query processor for a search language a {\em search processor\/}.
From analyzing popular search engines, one can define a set of criteria 
that should guide the designing of a search language and processor. 
We present such criteria below.
\begin{enumerate}
\item {\bf Format of Results:} A search result of a search query
should be either a set of documents (pages) or sections of documents
that satisfy the query.  In general, when searching, the user is
simply interested in {\em finding\/} information.  Thus, a search
query need not perform restructuring of documents to compute results.
This simplifies the formulation of a search query since the format of
the result need not be specified. \label{Criterion-Format}
\item {\bf Pattern Matching:} A search language should allow some level of pattern
matching both on the data and meta-data. Clearly, pattern matching on the data
is a convenient way of specifying search requirements. Pattern matching on the 
meta-data allows a user to formulate a search query without knowing the exact
structure of the document. In the context of searching, it is unlikely
that the user will be aware of the exact structure of the document that she is
seeking.\label{Criterion-Pattern:Matching}
\item {\bf Quantification:} Many search languages currently
implemented on the Web allow the user to specify quantifications in
search queries. For example, the search query ``+~Wild
-~West'', according to the semantics of many of the search engines
found on the Web, requests documents in which the word ``Wild''
appears (i.e., exists) and the word ``West'' does not appear (i.e.,
not exists). The ability to specify quantifications should be extended
to allow quantifications in querying the
meta-data. \label{Criterion-Quantification}
\item {\bf Logical Expressions:} Many search engines allow the user to
specify logical expressions in their search languages, such as
conjunctions and disjunctions of conditions. This should be extended
to enable the user to use logical expressions in querying the
meta-data. \label{Criterion-Logical:Expressions}
\item {\bf Iterative Searching Ability:} The result of a search query
is generally very large. Many times a result may contain hundreds, if
not thousands, of documents. Users generally do not wish to sift
through many documents in order to find the information that they
require. Thus, it is a useful feature for a search processor to allow
requerying of previous results. This enables users to search for the
desired information iteratively, until such information is
found. \label{Criterion-Requeying}
\item {\bf Polynomial Time:} The database over which search queries
are computed is large and is constantly growing. Hence, it is
desirable for a search query to be computable in polynomial time under
combined complexity (i.e., when both the query and the database are
part of the input). \label{Criterion-Polynomial}
\end{enumerate}

When designing a search language,
there is an additional requirement that is more difficult to define
scientifically. A search language should be {\em easy to use}. We present
our final criterion. 

\begin{enumerate}
\setcounter{enumi}{6}
\item {\bf Simplicity:} A search should be simple to use. One should be 
able to formulate queries easily and the queries, once formulated,
should be intuitively understandable. \label{Criterion-Simplicity}
\end{enumerate}

The definition of requirements for a search language is interesting in itself. 
In this paper we present a specific language, namely EquiX, that fulfills the
requirements~\ref{Criterion-Format} through~\ref{Criterion-Polynomial}.  From
our experience, we have found EquiX search queries to be intuitively
understandable.  Thus, we believe that EquiX satisfies the additional
language requirement of simplicity. 
EquiX is rather unique in that it combines both polynomial query evaluation 
(under combined complexity) with several powerful querying abilities. In 
EquiX, both quantification and negation can be used. In an extension to 
EquiX we allow aggregation and a limited class of regular expressions. 
Both searching and querying can be performed using the EquiX language. 
EquiX also simplifies the querying process by automatically 
generating both the format of the result and a corresponding DTD. 

This paper extends  previous work\cite{EquiX:Web-DB}.
In Section~\ref{Section-Data:Model} we present a data model for XML documents.
Both the concrete and abstract syntax for EquiX queries are described in 
Section~\ref{Section-Syntax}. In Section~\ref{Section-Semantics}
we define the semantics of EquiX, and in 
Section~\ref{Section-Query:Evaluation} a polynomial algorithm for evaluating
EquiX queries is presented. 
In Section~\ref{Section-Extending:EquiX} we
present some extensions to our language and in Section~\ref{Section-Conclusion}
we conclude. 
A procedure for computing a result DTD is presented
in Appendix~\ref{Section-Creating:Result}. We prove the correctness of our
evaluation algorithm in Appendix~\ref{Section-Proofs}.


\section{Data Model}
\label{Section-Data:Model}

We define a data model for querying XML documents~\cite{XML}. 
At first, we assume that each XML document has a given DTD. In 
Section~\ref{Section-Extending:EquiX}
we will relax this assumption. The term {\em element\/} will
be used to refer to a particular occurrence of an element in a 
document. The term {\em element name\/} will refer to 
a name of an element and thus, may appear many times in a document.
Similarly we use {\em attribute\/} to refer to a particular occurrence of an 
attribute and {\em attribute name\/} to refer to its name.
At times, we will blur the distinction between these terms when the meaning is
clear from the context. 

We introduce some necessary notation. 
A {\em directed tree\/} over a set of nodes $N$ is a 
pair $T = (N,E)$ where 
$E\subseteq N\times N$ and $E$ defines a tree-structure. 
We say that the edge $(n,n')$ is {\em incident from\/}
$n$ and {\em incident to\/} $n'$. Note that in a tree, there is at most
one edge incident to any given node. We assume throughout this paper that 
all trees are finite.
A directed tree is {\em rooted\/} if there is a 
designated
node $r\in N$, such that every node in $N$ is reachable from $r$ in $T$.
We call $r$ the {\em root\/} of $T$. We denote a rooted directed tree as a triple
$T = (N,E,r)$.

An XML document contains both data (i.e., atomic values) and meta-data (i.e., 
elements and attributes). 
The relationships between data and meta-data, 
(and between meta-data and meta-data) are reflected in a document 
by use of nesting.

We will represent an XML document by a directed tree with a labeling function.
The data and meta-data in a
 document correspond to nodes in the tree with appropriate labels. Nodes
corresponding to meta-data are {\em complex nodes\/} while
nodes corresponding to data are {\em atomic nodes\/}. The
relationships  in a document are represented by
edges in the tree. 
In this fashion, an XML document is represented by its 
parse tree.

Note that using {\tt ID} and {\tt IDREF} attributes one can represent
additional relationships between values. When considering these relationships, 
a document may no longer be represented by a tree. In the sequel we will utilize
{\tt ID} and {\tt IDREF} attributes to answer search queries.

In general, a parsed XML document need not be a rooted
tree. An XML document that gives rise to a rooted tree is said to be 
{\em rooted\/}. The element that corresponds to the root of the tree is
called the {\em root element}. 
Given an XML document that is not rooted, one can create a rooted document 
by adding a new element to the document and placing its 
opening tag at the beginning of the document, and its closing tag at the end
of the document. This new element will be the root element of the new document.
With little effort we can adjust the DTD of the original document to create 
a new DTD that the new document will conform to. Thus, we assume without loss
of generality that all XML documents in the database are rooted. 

We now give a formal definition of an XML document. We assume that there is 
an infinite set $\A$ of atoms and infinite set $\L$ of labels. 
\begin{definition}[XML Document]
An {\em XML document\/} is a pair $X = (T,l)$ such that 
\begin{itemize}
	\item $T = (N,E,r)$ is a rooted directed tree\footnote{Note that 
an XML document is a sequence of characters. Thus, to properly model the ordering 
of elements in a document, an ordering function on the children of a node should 
be introduced. For simplicity of exposition we chose to omit this in the paper.};
	\item $l:N\rightarrow \L\cup\A$ is a labeling function that
associates each complex node with a value in $\L$ and each atomic node
with a value in $\A$.
\end{itemize}
\end{definition}

We assume that each DTD has a designated element name, called the 
{\em root element name\/} of the DTD. Consider a DTD $d$ with a root 
element name $e$. We say that a document $X = (T,l)$ with root $r$ 
{\em strictly conforms to\/} $d$ if 
\begin{enumerate}
\item the document $X$ conforms to $d$ (in the usual way~\cite{XML}) and 
\item the function $l$ assigns the label $e$ to the root $r$ 
(i.e., $l(r) = e$).
\end{enumerate}

The following DTD with root element name {\tt movieInfo} describes information
 about  movies.
\noindent {\tt \begin{verbatim} 
<!ELEMENT movieInfo   (movie+,actor+)>
<!ELEMENT movie       (descr,title,character+)>
<!ELEMENT actor       (name)>
<!ATTLIST actor
          id          ID      #REQUIRED>
<!ELEMENT descr       (#PCDATA)>
<!ELEMENT title       (#PCDATA)>
<!ELEMENT name        (#PCDATA)>
<!ELEMENT character   EMPTY>
<!ATTLIST character
          role        CDATA   #REQUIRED
          star        IDREF   #REQUIRED>
\end{verbatim}
}
In Figure~\ref{Figure-Document}  an XML document
 containing movie information is depicted. 
This document strictly conforms to the DTD resented above.

\sizefig{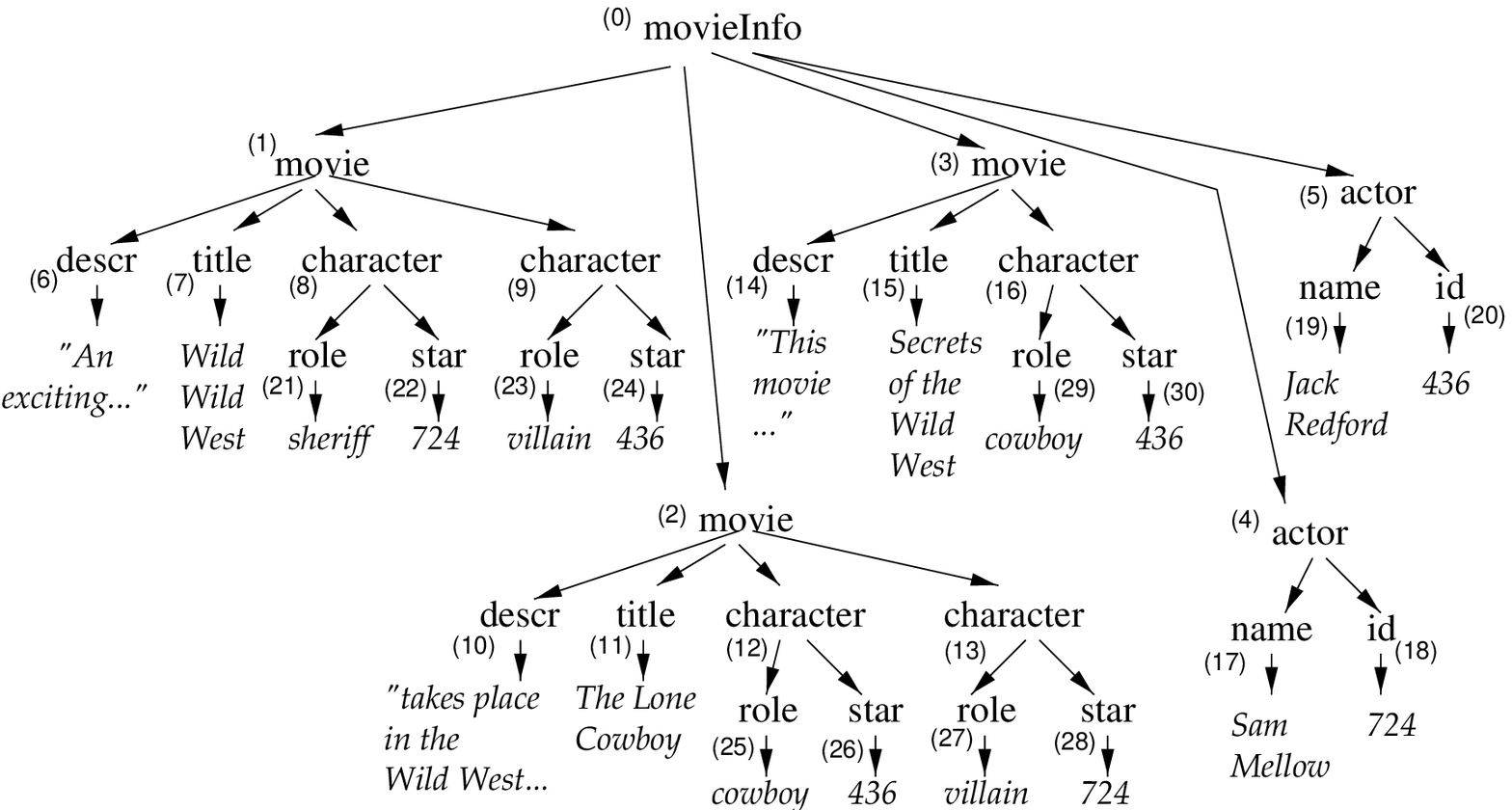}{0.8\textwidth}{An XML Document}{Figure-Document}
Note that the nodes in Figure~\ref{Figure-Document} are numbered. The numbering
is for convenient reference and is not part of the data model.

A {\em catalog\/} is a pair $C = (d, S)$ where $d$ is a DTD and $S$ is
a set of XML documents, each of which strictly conforms to $d$. A {\em
database\/} is a set of catalogs. Note the similarity of this
definition to the relational model where a database is a set of tuples
conforming to given relation schemes.

This data model is natural and has useful characteristics. 
Our assumption that each XML document conforms to a given DTD
implies that the documents are of a partially known structure.
We can display this knowledge for the benefit of the user. 
Thus, the task of finding information in a database
does not require a preliminary step of querying the database 
to discover its structure.

\section{Search Query Syntax}
\label{Section-Syntax}

In this section we present both a concrete and an abstract syntax for EquiX 
search queries. A search query written in the concrete syntax is a {\em concrete query\/}
and a search query written in the abstract syntax is an {\em abstract query\/}.

\subsection{Concrete Query Syntax}
The concrete syntax is described informally as part of the graphical
user interface currently implemented for EquiX. 
Intuitively, a query is an ``example'' of 
the documents that should appear in the output. 
By formulating an EquiX query the user can specify documents that she would 
like to find. She can specify constraints on the data that should appear in 
the documents. We call such constraints {\em content constraints\/}. 
She can also specify constraints on the meta-data, or structure, of the 
documents. We call such constraints {\em structural constraints\/}.
In addition, the user can specify {\em quantification constraints\/} which 
constrain the data and meta-data that should appear in the resulting
documents by determining how the content and structural constraints should
be applied to a document.

The user formulates her query interactively. The user chooses a catalog 
$(d,S)$. 
Only documents in $S$ will be searched (queried).
At first a {\em minimal query\/} is 
displayed. In a minimal query, only the root element name of $d$ is displayed.
A minimal query looks similar to an empty form for querying using a 
search engine (see Figure~\ref{Figure-Minimal:Query}). The user can then 
add content constraints by filling in the form, or add structural constraints
by expanding elements that are displayed. When an 
element is expanded, its attributes and subelements, as defined in $d$,
are displayed. The user can add content constraints to the elements and attributes.
The user can also specify the quantification
that should be applied to each element and attribute, i.e., quantification constraints. 
This can be one of {\em exists\/}, {\em not exists\/}, {\em for all\/}, and
{\em not for all\/} (written in a user friendly fashion).
In addition, the user can choose which elements
in the query should appear in the output. 
\sizefig{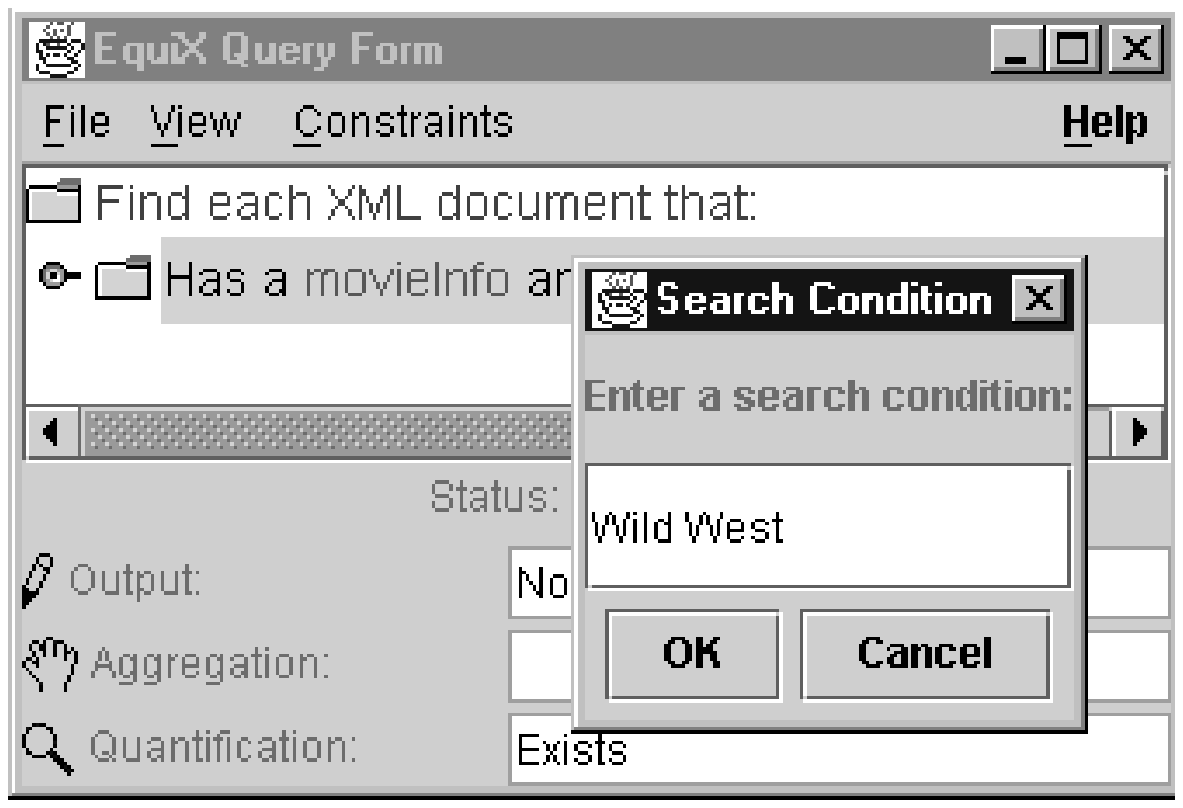}{0.6\textwidth}{Minimal query that finds documents containing the phrase ``Wild West''}{Figure-Minimal:Query}

In Figure~\ref{Figure-Concrete:Query} an expanded concrete query is depicted. 
This query was formulated by exploring the DTD presented in 
Section~\ref{Section-Data:Model}. It retrieves the title and description 
of Wild West movies in which Redford does not star as a villain. Intuitively,
answering this query is a two part process:
\begin{enumerate}
	\item {\em Search\/} for Wild West movies. The phrase ``Wild West'' 
may appear anywhere in the description of a movie. For example, it may appear
in the title or in the movie description. Intuitively, this is similar to a 
search in a search engine.
	\item {\em Query\/} the movies to find those in which Redford does not
play as a villain. This condition is rather exact. It specifies exactly 
where the phrases should appear and it contains a quantification constraint. 
Thus, conceptually, this is similar to a traditional database query.
\end{enumerate}

\sizefig{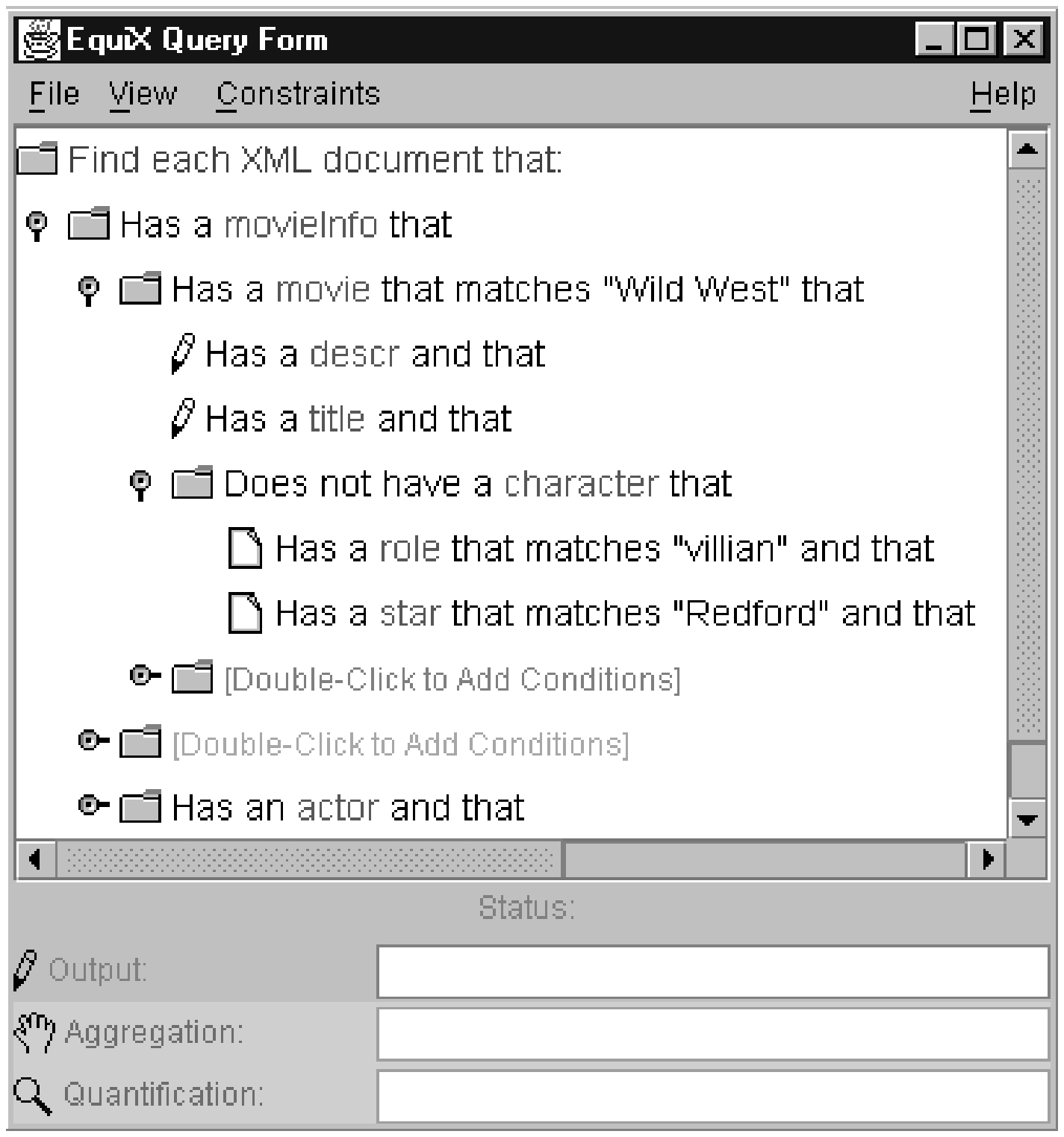}{0.6\textwidth}{Query that finds titles and descriptions of movies in which Redford isn't a villain}{Figure-Concrete:Query}
\subsection{Abstract Query Syntax}
We will present an abstract syntax for EquiX and show how a concrete query 
is translated to an abstract query.

A boolean function that associates each sequence
 of alpha-numeric symbols with a truth value among $\set{\bot,\top}$ is a
{\em string matching function\/}.
We assume that there is an infinite set $\C$ of string matching functions,
that $\C$ is closed under complement and that the function $\top$ is a
member of $\C$.
One such function might be:
	\[c_{wild\wedge west}(s) = \left\{ \begin{array}{ll}
			\top & \mbox{if $s$ contains the words wild and west}\\
			\bot & \mbox{otherwise}
			\end{array}
	\right. \]
We define an abstract query below.

\begin{definition}[Abstract Query]
An abstract query is a rooted directed tree $T$ augmented by four 
constraining  functions and an output set, denoted 
$Q = (T,l,c,\qao,\qef, O)$ where
\begin{itemize}
	\item $l:N\rightarrow \L$ is a {\em labeling function\/} that 
associates each node with a label;
	\item $c:N\rightarrow \C$ is a {\em content function\/}
 that associates each node with a string matching function;
	\item $\qao:N\rightarrow \set{\wedge, \vee}$ is an {\em operator
 function\/} that associates each node with a logical operator;
	\item $\qef:E\rightarrow \set{\exists, \forall}$ is a {\em 
quantification function\/} that associates each edge with a
quantifier;
	\item $O\subseteq N$ is the set of {\em projected nodes\/}, i.e.,
nodes that should appear in the result.
\end{itemize}
\end{definition}
Consider a node $n$. If $\qao(n) = \wedge$, we will say that $n$ is an
{\em and-node\/}.  Otherwise we will say that $n$ is an {\em
or-node\/}. Similarly, consider an edge $e$.  If $\qef(e) = \exists$,
we will say that $e$ is an {\em existential-edge\/}. Otherwise, $e$ is
a {\em universal-edge\/}.

We give an intuitive explanation of the meaning of an abstract
query. The formal semantics is presented in
Section~\ref{Section-Semantics}.  When evaluating a query, we will
attempt to {\em match\/} nodes in a document to nodes in the query. In
order for a document node $n_X$ to match a query node $n_Q$, the
function $c(n_Q)$ should hold on the data below $n_X$. In addition, if
$n_Q$ is an and-node (or-node), we require that each (at least one)
child of $n_Q$ be matched to a child of $n_X$.  If $n_X$ is matched to
$n_Q$ then a child $n'_X$ of $n_X$ can be matched to a child $n'_Q$ of
$n_Q$, only if the edge $(n_Q,n'_Q)$ can be {\em satisfied\/} w.r.t.\
$n_X$.  Roughly speaking, in order for an universal-edge
(existential-edge) to be satisfied w.r.t.\ $n_X$, all children (at
least one child) of $n_X$ that have the same label as $n'_Q$ must be
matched to $n'_Q$.

Note that in a concrete query the user can use the quantifiers 
$\set{\exists,\forall,\neg\exists,\neg\forall}$ and all nodes are implicitly
and-nodes. In an abstract query only the quantifiers $\set{\exists,\forall}$
 may be used and the nodes may be either and-nodes or or-nodes.
When creating a user interface for our language we found that the 
concrete query language was generally more intuitive for the user. We
present the abstract query language to simplify the discussion of the semantics
and query evaluation. Note that the two languages are equivalent in their
expressive power.

We address the problem of translating a concrete query to an abstract 
query. Most of this process is straightforward. The tree structure of the
abstract query is determined by the structure of the concrete query. The 
labeling function $l$ is determined by the labels (i.e., element and attribute
 names) 
appearing in the concrete query. 
The set $O$ is determined by the nodes marked for output by the user. 

Translating the quantification constraints is slightly more complicated. 
As a first 
step we augment each edge in the query with the appropriate quantifier as
determined by the user. We associate each node with the $\wedge$ operator
and with the content constraint specified by the user. Note that an empty
content constraint in a concrete query corresponds to the boolean function
$\top$.
Next, we propagate the negation in the query. When negation is propagated 
through an and-node (or-node), the node becomes an or-node (and-node), 
and the string matching function associated with the node is replaced by 
its complement.
Similarly, when negation is propagated through an existential-edge 
(universal-edge), the edge becomes a universal-edge (existential-edge).
In this fashion, we derive a tree in which 
each edge is associated with  $\exists$ or $\forall$ and each node is 
associated with $\wedge$ or $\vee$.
The functions $\qao$, $\qef$, and $c$ are determined by the process described
above. 
 
The concrete query in 
Figure~\ref{Figure-Concrete:Query} is represented by the abstract query in 
Figure~\ref{Figure-Abstract:Query}.  The string matching functions are
specified in italics next to the corresponding nodes. Black nodes are output
nodes. In the sequel, unless otherwise specified, the term {\em query\/} will
refer to an abstract query.
\sizefig{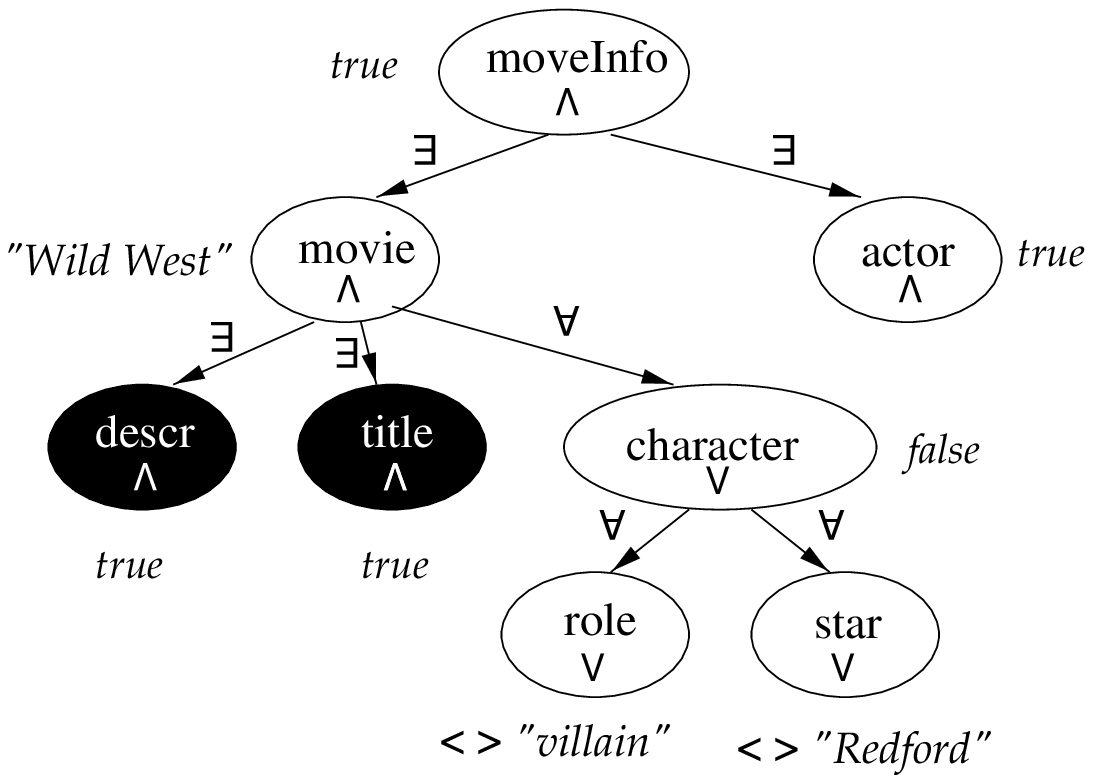}{0.6\textwidth}{Abstract query for the concrete query in Figure~\ref{Figure-Concrete:Query}. Output nodes are colored black.}{Figure-Abstract:Query}

Recall the search language requirements we presented in Section~\ref{Section-Introduction}.
We postulated that in a search language, it should not be necessary for the user to specify
format of the result (Criterion~\ref{Criterion-Format}). In EquiX, by defining the set $O$,
the user only specifies what information should she
wants the result to include, and does not explicitly detail the format in which it should 
appear. We suggested that it is important for there to be pattern 
matching,
quantification, and logical expressions for constraining data and meta-data 
(Criterion~\ref{Criterion-Pattern:Matching}, \ref{Criterion-Quantification}, 
and~\ref{Criterion-Logical:Expressions}). For data, these
can all be specified using the content function $c$. For meta-data, the pattern to which the
structure should be matched is specified by  $T$ and $l$, the quantification is specified by $\qef$, and logical operators can be specified using $\qao$. The result of an EquiX
query is a set of XML documents. In Section~\ref{Section-Creating:Result} we show
how a DTD for the result documents can be computed. Thus, requerying of results is possible
in EquiX (Criterion~\ref{Criterion-Requeying}). In 
Section~\ref{Section-Query:Evaluation} we show that EquiX queries can be 
evaluated in polynomial time, and thus, EquiX meets Criterion~\ref{Criterion-Polynomial}.

\section{Search Query Semantics} \label{Section-Semantics}

When describing the semantics of a query in a relational database language, 
such as SQL or Datalog, the term {\em matching\/} can be used. The result
of evaluating a query are all the tuples that match the schemas mentioned in the query
and satisfy the constraints. We describe the semantics of an EquiX query
in a similar fashion.

We first define when a node in a document matches a node in a query.
Consider a document $X$, and a query $Q$.  Suppose that the labeling
function of $X$ is $l_X$ and the labeling function of $Q$ is
$l_Q$.  We say that a node $n_X$ in $X$ {\em matches\/} a node
$n_Q$ in $Q$ if $l_X(n_X) = l_Q(n_Q)$.  We denote the parent
of a node $n$ by $p(n)$. We now define a matching of a document to a
query.

\begin{definition}[Matching] \label{Definition-Matching}
Let  $X=(T_X, l_X)$ be an XML document, with nodes $N_X$ and root $r_X$,
Let $Q=(T_Q,l_Q,c,\qao,\qef,O)$ be a query tree with nodes $N_Q$ and root $r_Q$.
A {\em matching of $X$ to $Q$\/} is a function
$\mu:N_Q\rightarrow 2^{N_X}$, such that the following hold
\begin{enumerate}
\item {\bf Roots Match:} $\mu(r_Q) = \{r_X\}$; 
	\label{Matching:Roots}
\item {\bf Node Matching:} if $n_X\in \mu(n_Q)$, $n_X$ matches $n_Q$;
	\label{Matching:Label}
\item {\bf Connectivity:} if $n_X\in \mu(n_Q)$ and $n_X$ is not the
	root of $X$, then $p(n_X)\in \mu(p(n_Q))$.
	\label{Matching:Connected}
\end{enumerate}
\end{definition}
Note that Condition~\ref{Matching:Roots} requires that the root of the document
is matched to the root of the query,  Condition~\ref{Matching:Label} insures
that matching nodes have the same label, and Condition~\ref{Matching:Connected}
requires matchings to have a tree-like structure. 

We define when a matching of a document to a query is satisfying. 
We first present some auxiliary definitions.
Consider an XML document $X = (T_X,l_X)$, where $T_X = \GraphOf X$. 
Consider a node $n_X$ in $T_X$. 
We differentiate between the {\em textual content\/} (i.e., data) 
contained below the node $n_X$, and the structural content (i.e., meta-data).
 When defining the textual content of a node,
we take {\tt ID} and {\tt IDREF} values into consideration. We say that $n'_X$ is a 
{\em child\/} of $n_X$ if $(n_X,n'_X)\in E_X$. We say that $n'_X$ is an
{\em indirect child\/} of $n_X$ if $n_X$ is an attribute of type {\tt IDREF} with
the same value as an attribute of type {\tt ID} of $n'_X$.
We denote the textual content of a node $n_X$ as $t(n_X)$, defined
\begin{itemize}
	\item If $n_X$ is an atomic node, then $t(n_X)=l_X(n_X)$;
	\item Otherwise, $t(n_X)$ is a
concatenation\footnote{%
Note that an XML document may be cyclic as a result of ID and IDREF
attributes. We take a finite concatenation by taking each child into
account only once. In addition, the order in which the concatenation is taken
and the ability to differentiate between data that originated in different
nodes may affect the satisfiability of a string matching function. This is
technical problem that is taken into consideration in the implementation.
We will not elaborate on this point any further. } of the content of its children and indirect
children.
\end{itemize}

We demonstrate the textual content of a node with an example. 
Recall the XML document depicted in Figure~\ref{Figure-Document}. 
The textual content of Node 9, is ``villain 436 Jack Robinson''.
Note that the $t(24)$ includes the value ``Jack Robinson'' since Node 
5 is an indirect child of Node 24.

We discuss when a quantification constraint is satisfied.
Consider a document $X$, a query $Q$ and a matching  $\mu$ of $X$ to $Q$.
 Let $n_X$ be a node in $X$ and let $e = (n_Q,n'_Q)$ be an edge in 
$Q$.
We say {\em $n_X$ satisfies $e$ with respect to $\mu$\/} if the following
holds
\begin{itemize}
	\item If $e$ is an existential-edge then there is a child $n'_X$ of
$n_X$ such that $n'_X$ matches $n'_Q$ and $n'_X\in\mu(n'_Q)$.
	\item If $e$ is a universal-edge then for all children $n'_X$ of $n_X$,
if $n'_X$ matches $n'_Q$, then $n'_X\in\mu(n'_Q)$.
\end{itemize}

We define a satisfying matching of a document to a query.

\begin{definition}[Satisfying Matching] \label{Definition-Satisfying:Matching}
Let  $X=(T_X, l_X)$ be an XML document, 
and let $Q=(T_Q,l_Q,c,\qao,\qef,O)$ be a query tree.
Let $\mu$ be a matching of $X$ to $Q$. We say that 
$\mu$ is a {\em satisfying mapping of $X$ to $Q$\/} if 
for all nodes $n_Q$ in $Q$ and for all nodes $n_X\in\mu(n_Q)$ 
the following conditions hold
\begin{enumerate}
	\item if $n_Q$ is a leaf then $c(n_Q)(t(n_X)) = \top$, i.e., 
$n_X$ satisfies the string matching condition of $n_Q$;
\label{def:satisfying:matching:content}
	\item otherwise ($n_Q$ is not a leaf):
	\begin{enumerate}
	\item if $n_Q$ is an or-node then $n_X$ satisfies either $c(n_Q)$ 
or at least one edge incident from $n_Q$ with respect to $\mu$;
	\label{def:satisfying:matching:or}
	\item if $n_Q$ is an and-node then $n'_X$ satisfies both $c(n_Q)$ 
and all edges that are incident from $n_Q$ with respect to $\mu$.
	\label{def:satisfying:matching:and}
	\end{enumerate}
\end{enumerate}
\end{definition}
Condition~\ref{def:satisfying:matching:content} implies that the leaves 
satisfy
the content constraints in $Q$. Conditions~\ref{def:satisfying:matching:or}
and~\ref{def:satisfying:matching:and} imply that $X$ satisfies the 
quantification constraints in $Q$. The structural constraints are satisfied 
by the existence of a matching. 

\begin{example}
Recall the query in Figure~\ref{Figure-Abstract:Query} and the document in 
Figure~\ref{Figure-Document}. Two satisfying matchings of the document to the
query are specified in the following table. There are additional 
matchings not shown here.
\begin{center}
\begin{tabular}{|l||l|l|} \hline
{\em Query Node} & $\mu_1$           &  $\mu_2$       \\ \hline \hline
movieInfo        & $\set{0}$         &  $\set{0}$     \\ \hline
movie            & $\set{2}$         &  $\set{3}$     \\ \hline
descr            & $\set{10}$        &  $\set{14}$    \\ \hline
title            & $\set{11}$        &  $\set{15}$    \\ \hline
character        & $\set{12,13}$     &  $\set{16}$    \\ \hline
role             & $\set{25,27}$     &  $\set{29}$    \\ \hline
star             & $\set{26,28}$     &  $\set{30}$    \\ \hline
actor            & $\set{4}$         &  $\set{5}$     \\ \hline
\end{tabular}
\end{center}
Note that there is no satisfying matching that matches Node 1 to the movie
node in the query because the universal quantification on the edge connecting
movie and character cannot be satisfied.
\end{example}

We presented several matchings of a document to a query.  Let $\mu$ and $\mu'$ be matchings of a 
document $X$ to a query $Q$. We define the {\em union\/} of $\mu$ and
$\mu'$ in the obvious way. Formally, given a query node $n_Q$, 
\[ (\mu \cup \mu')(n_Q) := \mu(n_Q) \cup \mu'(n_Q) \]

There may be an exponential number of matchings of a given document to a given 
query. Note, however, that the following proposition holds.

\begin{proposition}[Union of Matchings]
Let $X$ be an XML document and let $Q$ be a query. Let $\M$ be the set of
all matchings of $X$ to $Q$. Then the union of all the matchings in $\M$ is
a matching. Formally,
	\[ (\bigcup_{\mu\in\M} \mu)\in \M \]

\end{proposition}

We say that a document $X$ {\em satisfies\/} a query $Q$ if there exists
a satisfying matching $\mu$ of $X$ to $Q$. 
We now specify the output of evaluating a query on a single XML document.
The result of a query is the set of documents derived by evaluating  the query 
on each document in the queried catalog (i.e., each document that matched the
DTD from which the query was derived).

Intuitively, the result of evaluating a query on a document is
a subtree of the document (as required in Criterion~\ref{Criterion-Format}). 
The subtree contains nodes of three types. Document nodes corresponding to 
{\em output\/} query nodes appear in the resulting subtree. In addition, 
we include {\em ancestors\/} and {\em descendents\/} of these nodes. The
ancestors insure that the result has a tree-like structure and that it
is a projection of the original document. 
Recall that the textual content of the document is contained in the atomic
nodes of
the document tree. Hence, the result must include the descendents to insure
that the the textual content is returned.

For a given document, query processing can be viewed as the process of
singling out the nodes of the document tree that will be part of the output. 
Consider a document $X=(T_X,l_X)$ with $T_X = \GraphOf X$ and a query $Q$ with
projected nodes $O$. 
Let $\M$ be the set of satisfying matchings of $X$ to $Q$.
The output of evaluating the query $Q$ on the document $X$ is the
the document defined by projecting $N_X$ on the set 
$N_R:=N_{\mbox{out}}\cup N_{\mbox{anc}} \cup N_{\mbox{desc}}$ defined as
\begin{itemize}
\item 
$N_{\mbox{out}}:=
 \set{n_X \in N_X\mid (\exists n_O\in O)(\exists \mu\in\M)\ n_X\in\mu(n_O)}$, i.e.,
nodes in $X$ corresponding to projected nodes in $Q$;
\item 
$N_{\mbox{anc}}:=
	\set{n_X\in N_X\mid (\exists n'_X\in N_{\mbox{out}}) 
	\mbox{ $n_X$ is an ancestor of $n'_X$}}$, i.e., ancestors of nodes in $N_{\mbox{out}}$;
\item 
$N_{\mbox{desc}}:=\set{n_X\in N_X\mid (\exists n'_X\in  N_{\mbox{out}}) 
	\mbox{ $n_X$ is an descendent of $n'_X$}}$, i.e.,
descendents of nodes in $N_{\mbox{out}}$.
\end{itemize}
We call $N_R$ the {\em  output set\/} of $X$ with respect to $Q$.

The result of applying the query in Figure~\ref{Figure-Abstract:Query} to 
the document in Figure~\ref{Figure-Document} is depicted in 
Figure~\ref{Figure-Result}. Note that the values of ``descr'' and ``title'' are
grouped by ``movie''. This follows naturally from the structure of the original
document.
\sizefig{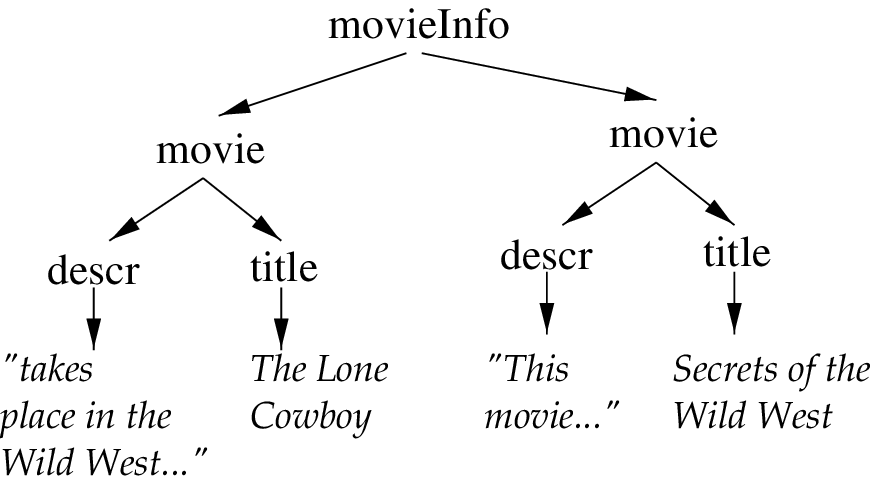}{0.4\textwidth}{Result Document}{Figure-Result}
\section{Query Evaluation} \label{Section-Query:Evaluation}

Recall that a query is defined by choosing a catalog and exploring its DTD. 
Consider a query $Q$ generated from a DTD $d$ in the catalog $(d,S)$. 
The result of 
evaluating $Q$ on the database is the set of documents generated by 
evaluating $Q$ on each document in $S$.

We present an algorithm for evaluating a query on a document.
There may be an exponential number of matchings of a query to a document.
Concrete queries contain both quantification and negation.  This would
appear to be another source of complexity. Thus, it
would seem that computing the output of a query on a document should
be computationally expensive. 
Roughly speaking, however, query evaluation in this case
is analogous to evaluating a first-order query that can be written
using only two variables. Therefore, using dynamic 
programming\cite{Cormen:Algorithms} we can
in fact derive an algorithm that runs in polynomial time, even when
the query is considered part of the input (i.e., combined complexity).
Thus, EquiX meets the search language requirement of having polynomial
 evaluation time (Criterion~\ref{Criterion-Polynomial}).

In Figure~\ref{Figure:Evaluation} we present a polynomial procedure that 
computes the output of a document, given a query. Given a document
$X$ and query $Q$, the procedure  {\sf Query\_Evaluate} computes  
the output set $N_R$ of $X$ w.r.t.\ $Q$.
Note that the value of $t(n_X)$ for each document node $n_X$ can be computed
in a preprocessing step polynomial time.
{\sf Query\_Evaluate}
 uses the procedure {\sf Matches} shown in Figure~\ref{Figure:Satisfy}.
Given a query node $n_Q$ and a document node $n_X$,
the procedure {\sf Matches} checks if it is possible that $n_X\in\mu(n_Q)$
for some matching $\mu$, based on the subtrees of $n_Q$ and $n_X$.

\newcommand{\nout}{N_{\mbox{out}}}

\begin{figure}[htb]
\begin{center}
\fbox{
\parbox{6in}{
\begin{tabbing}
AAAAA \= AAAAA\= AAAAA\= AAAAA\=\kill
ElseJ \= ElseJAA \= AAAA\= AAAAA\=\kill
{\bf Algorithm} \>\> {\sf Query\_Evaluate}\\
{\bf Input}     \>\> A document $X = (T_X,l_X)$ s.t.\ 
                        $T_X = \GraphOf X$, \\
                \>\> A query tree $Q= \queryOf Q$ s.t.\ $T_Q = \GraphOf Q$.\\
{\bf Output}    \>\> $N_R\subseteq N_X$, i.e., the outputed document nodes \\
\\
{\bf Initialize} {\it match\_array[][]} to {\bf false} \\
{\it Queue$_1$} $:= N_Q$, ordered by descending depth \\
{\bf While} ({\bf not} {\it isEmpty(Queue$_1$)}) {\bf do} \\
        \> $n_Q := {\it Dequeue(Queue_1)}$ \\
        \> {\bf For all} $n_X \in N_X$ such that ${\it path}(n_X) = 
                                        {\it path}(n_Q)$ {\bf do}\\
        \>      \> {\it match\_array}[$n_Q$,$n_X$]$:=$ {\sf Matches}($n_Q$,
                                                $n_X$,{\it match\_array}) \\
$N_R := \emptyset$ \\
{\it Queue$_2$} $:= N_Q$, ordered by ascending depth \\
{\bf While} ({\bf not} {\it isEmpty(Queue$_2$)}) {\bf do} \\
        \> $n_Q := {\it Dequeue(Queue_2)}$ \\
        \> {\bf For all} $n_X\in N_X$ {\bf do} \\
        \>      \> {\bf If} ($n_Q\neq r_Q$ {and \bf not} {\it match\_array}[$p(n_Q)$,$p(n_X)$]) {\bf then} \\
        \>\>\> {\it match\_array}[$n_Q$,$n_X$] := {\bf false}\\
        \>     \> {\bf Else If} ({\it match\_array}[$n_Q$,$n_X$] {\bf and}
                                $n_Q\in O$) {\bf then} \\
        \>      \>      \> $N_R:=N_R\cup \set{n_X}\cup{\it anc}(n_X)\cup {\it desc}(n_X)$\\
{\bf Return} $N_R$
\end{tabbing}
}}
\end{center}
\caption{Evaluation of an EquiX Query}
\label{Figure:Evaluation}
\end{figure}

\begin{figure}[htb]
\begin{center}
\fbox{
\parbox{10in}{
\begin{tabbing}
ElseJ \= ElseJ \= AAAA\= AAAAA\=\kill
{\bf Procedure} \>\>\> {\sf Matches}($n_Q$,$n_X$,{\it match\_array})\\
{\bf Input}     \>\>\> A query node $n_Q$ \\
                \>\>\> A document node $n_X$ \\
                \>\>\> An array {\it match\_array} \\
{\bf Output}    \>\>\> {\bf true} if $n_X$ may be in $\mu(n_Q)$ for a matching $\mu$,\\
                \>\>\>  based on the subtrees of $n_X$ and $n_Q$, and {\bf false}
 otherwise \\
\\
{\it tc} := $c(n_Q)(t(n_X))$ \\
{\bf If} $n_Q$ is a terminal node {\bf return} {\it tc} \\
{\bf Let} $M_Q$ be the set of children of $n_Q$ in $Q$ \\
{\bf For each} $m_Q\in M_Q$ do: \\ 
        \> {\bf Let} $M_X$ be the set of children, $m_X$, of $n_X$ in $X$
                         such that $l_X(m_X)=l_Q(m_Q)$ \\ 
        \> {\bf If} $(n_Q,m_Q)$ is an existential-edge {\bf then}\\
        \>      \> {\it status}$({m_Q})$ $:=$ \ \ $\bigvee_{m_X\in M_X}$
        {\it match\_array}[$m_Q$,$m_X$] \\
        \> {\bf Else}  \>{\it status}$({m_Q})$ $:=$ \ \ $\bigwedge_{m_X\in M_X}$ 
                   {\it match\_array}[$m_Q$,$m_X$] \\ 
{\bf If} $n_Q$ is an or-node {\bf then} \\
        \> {\bf return} {\it tc}$\vee (\bigvee_{m_Q\in M_Q}$ {\it status}$({m_Q}))$\\ 
{\bf Else} \> {\bf return} {\it tc}$\wedge(\bigwedge_{m_Q\in M_Q}$ {\it status}$({m_Q}))$  
\end{tabbing}
}}
\end{center}
\caption{Satisfaction of a Node Procedure}
\label{Figure:Satisfy}
\end{figure}

Note that {\it path}$(n)$ is the sequence of element names on the path
from the root of the query to $n$, and {\it anc($n$)} ({\it desc($n$)}) 
is the set of ancestors (descendents) of $n$.
 Note also that {\it match\_array\/}
is an array of size $N_Q\times N_X$ of boolean values, where $N_Q$ are
the nodes in the query and $N_X$ are the nodes in the document. 
Observe that in
Figure~\ref{Figure:Evaluation} we order the nodes by descending
depth. This insures that when {\sf Matches}($n_Q$,$n_X$,{\it
match\_array}) is called, the array {\it match\_array} is already
updated for all the children of $n_Q$ and $n_X$. The procedure {\sf
Query\_Evaluate} does not explicitly create any matchings.  However,
the following theorem holds.

\begin{theorem}[Correctness of {\sf Query\_Evaluate}]
Given document $X$ and a query $Q$, the algorithm {\sf Query\_Evaluate}
retrieves the output set of $X$ w.r.t.\ $Q$.
\end{theorem}

In Appendix~\ref{Section-Proofs} we prove this theorem. The 
procedure {\sf Query\_Evaluate} runs in polynomial time in combined complexity.
Let $|D|$ be the size of the data in document $X$, i.e., the size
of $X$ when ignoring $X$'s meta-data. Let $C(m)$ be an upper-bound
on the runtime of computing a string-matching constraint on a string of 
size $m$. 

\begin{theorem}[Polynomial Complexity]
Given document $X$ and a query $Q$, the algorithm {\sf Query\_Evaluate}
runs in time $O(|N_X|\cdot|N_Q|\cdot(|N_Q|\cdot|N_X| + C(|D|)))$.
\end{theorem}

Note that query evaluation generates a set of documents.
Recall that a query is formulated by exploring a DTD
and only documents in the catalog of the DTD chosen will be queried.
Thus, in order to allow 
{\em iterative querying\/} or {\em requerying of results\/}, a DTD
for the resulting documents must be defined. A {\em result DTD\/} is a DTD to
 which the resulting documents strictly conform. In 
Appendix~\ref{Section-Creating:Result} we present a polynomial procedure 
that computes a result DTD for a given query. The result DTD is linear in 
the size of the DTD from which the query was originated. The compactness of
the result DTD makes the requerying process simpler, since requerying entails
exploring the result DTD.
Thus, EquiX fulfills the search 
language requirement of ability to perform requerying 
(Criterion~\ref{Criterion-Requeying}).


\section{Extending EquiX Queries} \label{Section-Extending:EquiX}

EquiX can be extended in many ways to yield a more powerful
language. In this section we present two extensions to the EquiX
language. These extensions add additional querying ability to
EquiX. After extending EquiX, the search language
requirements~\ref{Criterion-Format} through~\ref{Criterion-Polynomial} 
are still met. However, it is a matter of opinion
if EquiX still fulfills Criterion~\ref{Criterion-Simplicity}
requiring simplicity of use.  Thus, these extensions are perhaps more
suitable for expert users.
 
\subsection{Adding Aggregation Functions and Constraints}
We extend the EquiX language to allow computing of aggregation
functions and verification of aggregation constraints. We call the new
language $\EquiXagg$.

We extend the abstract query formalism to allow aggregation. An {\em
aggregation function\/} is one of
$\set{\Count,\Min,\Max,\Sum,\Avg}$. An {\em atomic aggregation
constraint} has the form $f\theta v$ where $f$ is an aggregation
function, $\theta\in\set{<,\leq,=,\neq,\geq,>}$, and $v$ is a constant
value. An {\em aggregation constraint} is a (possibly empty)
conjunction of atomic aggregation constraints. In $\EquiXagg$ a query
is a tuple $\aggquery$ as in EquiX, augmented with the following
functions:
\begin{itemize}
	\item $a$ is an {\em aggregation specifying function\/} that
associates each node with a (possibly empty) set of aggregation
functions;
	\item $a_c$ is an {\em aggregation constraint function\/}
that associates each node with an aggregation constraint.
\end{itemize}
Given a node $n_Q$, the aggregation specifying functions $a(n_Q)$ 
(and similarly the
aggregation constraint functions) are applied to $t(n_Q)$.
Note that the $\Min$ and $\Max$ functions can only be applied to an argument
whose domain is ordered.
Similarly, the aggregation functions $\Sum$ and
$\Avg$ can only be applied to sets of numbers. There is no way to enforce 
such typing using a DTD (although it is possible using an XML 
Schema~\cite{XML-Schema}). 
When a function is applied to an argument that does 
not meet its requirement, its result is undefined. 

The function $a$ adds the computed aggregation values to the output. This
is similarly to placing an aggregation function in the SELECT clause of
an SQL query. The function $a_c$ is used to further constrain a query. This
is similar to the HAVING clause of an SQL query.

In order to use an aggregation function in an SQL query, one must include
a GROUP-BY clause. This clause specifies on which variables the grouping 
should be performed, when computing the result. To simplify the querying 
process, we do not require the user to specify the GROUP-BY variables. 
$\EquiXagg$
uses a simple heuristic rule to determine the grouping variables.
Suppose that $a(n_Q)\neq \emptyset$ for some node $n_Q$. Let $n_O$ be the 
lowest node above $n_Q$ in $Q$  for which one of the following conditions hold
\begin{itemize}
	\item $n_O\in O$ {\em or}
	\item $n_O$ is  an ancestor of a node in $O$.
\end{itemize}
Note that $n_O$ is the lowest node above $n_Q$ where both textual content and aggregate values may be combined.
$\EquiXagg$ groups by $n_O$ when
 computing  the aggregation functions on $n_Q$. In a similar fashion, 
$\EquiXagg$ performs grouping in order to compute aggregation constraints. 

Our choice for grouping variables is natural since it takes advantage
of the tree structure of the query, and thus, suggests a polynomial 
evaluation algorithm. It is easy to see that adding aggregation functions
and constraints does not affect the polynomiality of the evaluation algorithm.
The algorithm for creating a result DTD must also be slightly adapted 
in order to take into consideration the aggregation values that are retrieved.
Thus, $\EquiXagg$ meets the search language requirements~\ref{Criterion-Format}
through~\ref{Criterion-Polynomial}.

\subsection{Querying Ontologies using Regular Expressions}

In EquiX, the user chooses a catalog and queries only documents
in the chosen catalog. It is possible that the
user would like to query documents conforming to different DTDs, but
containing information about the same subject. In $\EquiXreg$ this
ability is given to the user. Thus, $\EquiXreg$ is useful for
information integration.

An ontology, denoted $\O$, is a set of terms whose meanings are well known.
Note that an ontology can be implemented using XML 
Namespaces~\cite{XML-Namespaces}.
We say that a document $X$ {\em can be described by\/} $\O$ if some
of the element and attribute names in $X$ appear in $\O$.
When formulating a query, the user chooses an ontology of terms that 
describes the subject matter that she is interested in querying. 
Documents that can be described by the chosen ontology will be queried. A 
query tree in $\EquiXreg$ is a tuple $\query$ as in EquiX. However,
in $\EquiXreg$, $l$ is a function from the set of nodes to $\O$.

Semanticly, an $\EquiXreg$ query is interpreted in a different fashion from
an EquiX query. Each edge is implicitly labeled with the ``+'' symbol. 
Intuitively, an edge in a query corresponds to a sequence of one or more
edges in a document. We adapt the definition of satisfaction of an edge
(presented in Section~\ref{Section-Semantics}) to reflect this change.

Consider a document $X$, a query $Q$ and a matching $\mu$ of $X$ to
 $Q$.  Let $n_X$ be a node in $X$ and let $e = (n_Q,n'_Q)$ be an edge
 in $Q$.  We say {\em $n_X$ satisfies $e$ with respect to $\mu$\/} if
 the following holds
\begin{itemize}
	\item If $e$ is an existential-edge then there is a 
{\em descendent\/} $n'_X$ of
$n_X$ such that $n'_X$ matches $n'_Q$ and $n'_X\in\mu(n'_Q)$.
	\item If $e$ is a universal-edge then for all {\em descendents\/}
 $n'_X$ of $n_X$, if $n'_X$ matches $n'_Q$, then $n'_X\in\mu(n'_Q)$.
\end{itemize}
Note that the only change in the definition was to replace the words child
and children with descendent and descendents.

In a straightforward fashion, we can modify the query evaluation algorithm
to reflect the new semantics presented. The new algorithm remains polynomial
under combined complexity. 
Note that it is no longer possible to create a result
DTD if we do not permit a result DTD to contain content definitions of type
{\tt ANY}. This results from the possible diversity of the documents being
queried. However, $\EquiXreg$ still meets Criterion~\ref{Criterion-Requeying}
(i.e., ability to requery results), since the resulting documents can 
be described by the chosen ontology. Thus,  $\EquiXreg$ meets the search 
language requirements~\ref{Criterion-Format} through~\ref{Criterion-Polynomial}.



\section{Conclusion}
\label{Section-Conclusion}

In this paper we presented design criteria for
a search language. We defined a specific search language for XML, 
namely EquiX, that fulfills these requirements. 
Both a user-friendly concrete syntax and a formal abstract syntax were
presented. We defined an evaluation algorithm for EquiX queries that is 
polynomial even under combined complexity. We also presented a polynomial
algorithm that generates a DTD for the result documents of an EquiX query. 

We believe that EquiX enables the user to search for information in a 
repository of XML documents in a simple and intuitive fashion. Thus, our
language is especially suitable for use in the context of the Internet.
EquiX has the ability to express
complex queries with negation, quantification and logical expressions.
We have also extended EquiX to allow aggregation and limited 
regular expressions.

Several XML query languages have been proposed recently, such as 
 XML-QL~ \cite{XML-QL}, 
XQL~\cite{XQL-Microsoft} and Lorel~\cite{xml-lorel:Web-DB}. 
These languages are powerful in their querying ability. However, they do
not fulfill some of our search language requirements. In these languages, the
user can perform restructuring of the result. Thus, the format of the result
must be specified, in contradiction to Criterion~\ref{Criterion-Format}. 
Furthermore, XML-QL and XQL are limited in their ability to express
quantification constraints (Criterion~\ref{Criterion-Quantification}). 
Most importantly, none of these languages guarantee polynomial evaluation
under combined complexity (Criterion~\ref{Criterion-Polynomial}). 

As future work, we plan to extend the ability of querying 
ontologies and to allow more complex regular expressions in EquiX.
XML documents represent data that may not have a strict schema.
In addition, search queries constitute a guess of the content of the desired
documents.
Thus, we plan to refine EquiX with the ability to deal with incomplete 
information 
\cite{Kanza:Nutt:Sagiv-Queries:with:Incomplete:Answers:over:SSD-PODS99} and
with documents that ``approximately satisfy'' a query.
Search engines
perform an important service for the user by sorting the results according
to their quality. We plan on experimenting to find a metric for ordering 
results that takes both the data and the meta-data into consideration.

As the World-Wide Web grows, it is becoming increasingly difficult for users
to find desired information. The addition of meta-data to the Web provides
the ability to both search and query the Web. Enabling users to formulate 
powerful queries in a simple fashion is an interesting and challenging problem.

\appendix
\section{Creating a Result DTD}
	\label{Section-Creating:Result}

In Section~\ref{Section-Query:Evaluation} we described the process
of evaluating a query on a database. 
Query evaluation generates a set of documents.
A query is formed using a chosen DTD, called the {\em originating DTD\/}, 
and only documents
strictly conforming to the originating DTD will be queried. 
Thus, in order to allow 
{\em iterative querying\/} or {\em requerying of results\/}, a DTD
for the resulting documents must be defined. Given a query $Q$, if
any possible result document must conform to the DTD $d_R$, we say that 
$d_R$ is a {\em result DTD for $Q$}.
In this section we
present a procedure that given a query $Q$, computes in polynomial time
a  result DTD for $Q$. Thus, we show that EquiX fulfills the search 
language requirement of ability to perform requerying (Criterion~\ref{Criterion-Requeying}).
	
A DTD is a set of {\em element definitions\/}, and {\em 
attribute list definitions\/}. An element definition has the form 
$<!{\tt ELEMENT}$~$e$ $\varphi>$, where $e$ is the element name being defined
and $\varphi$ is its {\em content definition\/}. 
An attribute list definition has
the form $<!{\tt ATTLIST}$~$e$ $\psi_1 \ldots \psi_n>$, where $e$ is an
element name and $\psi_1\ldots\psi_n$ are definitions of attributes for $e$.
The set of element names defined in a DTD $d$ is its {\em element name set\/}, 
denoted $\E_d$.

Consider a query $Q = \query$ formulated from a DTD $d$.
 We say that element name $e'$ is a 
{\em descendent\/} of element name $e$ in $d$ if $e'$ may be nested within an
element $e$ in a document conforming to $d$. Formally, $e'$ is a descendent
of $e$ if
\begin{itemize}
	\item $e'$ appears in the element definition of $e$ {\em or\/}
	\item $e'$ is a descendent of an element name $e''$ which appears in 
the content definition of $e$. 
\end{itemize}
We say that $e$ is an {\em ancestor\/} of $e'$ in $d$ if
$e'$ is a descendent of $e$ in $d$.
Note that element name $e$ may appear in 
a resulting document of evaluating $Q$ if
 there is a node $n_O\in O$ such that $l(n_O) = e$. Additionally,
$e$ may appear in the output if $e$ is an ancestor or descendent in $d$ of 
an element $e'$ that meets the condition presented in the previous sentence.
Thus, given a query, we can compute in linear time the element name set 
$\E_{d_R}$ of the result DTD $d_R$.

In order to compute the result DTD of a query $Q$, we must compute the 
content definitions and attribute list 
definitions for the elements in $\E_{d_R}$.
In the result DTD, we take the attribute list definitions for the elements in
$\E_{d_R}$ as defined in the originating DTD but change all attributes to be
of type {\tt \#IMPLIED}. Note that the root element name $r$ of the originating
DTD will always be in $\E_{d_R}$. It is easy to see that $r$ is also the designated
root element name $d_R$.

\begin{figure}[htb]
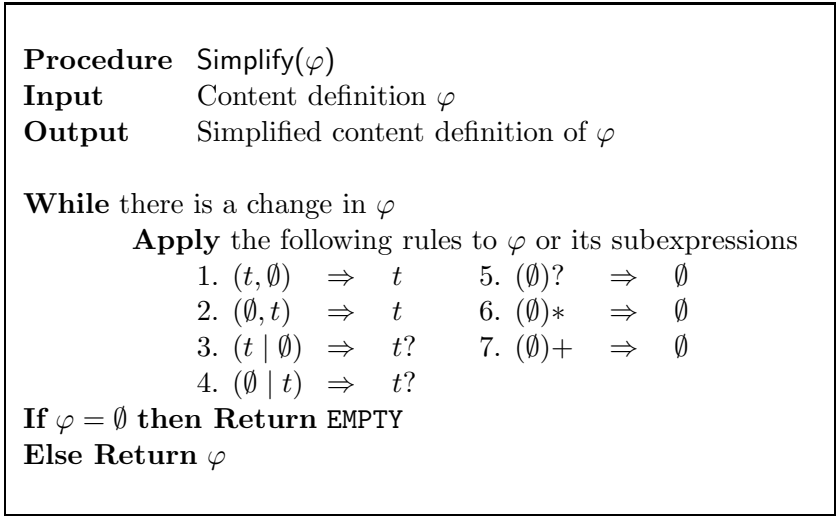

\begin{center}
\fbox{
\parbox{6in}{
\begin{tabbing}
AAAAA\=AAA\=AAAAAA\=AAA\=AAAA\=AAAAAA\=AAA\=AAAA\kill
{\bf Procedure} \>\> {\sf Simplify($\varphi$)}\\
{\bf Input}     \>\> Content definition $\varphi$ \ \ \ \ \\
{\bf Output}    \>\> Simplified content definition of $\varphi$\\
\\
{\bf While} there is a change in $\varphi$ \\
\> {\bf Apply} the following rules to $\varphi$ or its subexpressions\ \ \ \ \\
\> \>1. $(t, \emptyset)$ \> $\Rightarrow$ \> $t$ \> 
	5. $(\emptyset)?$ \> $\Rightarrow$ \> $\emptyset$  \\
\> \>2. $(\emptyset, t)$ \> $\Rightarrow$ \> $t$ \>
	6. $(\emptyset)*$ \> $\Rightarrow$ \> $\emptyset$ \\
\> \>3. $(t \mid \emptyset)$\> $\Rightarrow$ \> $t?$ \>
	7. $(\emptyset)+$ \> $\Rightarrow$ \> $\emptyset$ \\
\> \>4. $(\emptyset \mid t)$\> $\Rightarrow$ \> $t?$ \\
{\bf If} $\varphi = \emptyset$ {\bf then Return} {\tt EMPTY} \\
{\bf Else Return} $\varphi$
\end{tabbing}
}}
\end{center}
\caption{Definition Simplifying Algorithm}
\label{Simplify}
\end{figure}

In Figure~\ref{CDD} we present an algorithm that computes the content 
definition for an element in $\E_{d_R}$.
Intuitively, any elements that will not appear in
the result of a query must be removed from the original DTD in
order to form the result DTD. In addition, elements will only appear
in result documents if query constraints are satisfied. Thus, this
possible appearance of elements may be taken into account when formulating
$d_R$.
The algorithm {\sf Create\_Content\_Definition} uses the procedure presented
 in Figure~\ref{Simplify} 
in order to simplify the content definition it creates. The
result DTD  is created by computing the content definitions for all 
$e\in\E_{d_R}$
and adding the attribute list definitions. Note that in the algorithm,
$dtd\_desc(e',e)$ is true if $e$ is a descendent of $e'$ in the DTD $D$.
In addition, $anc(n_X,O)$ is true if $n_X$ is an ancestor of some node in 
$O$.

\begin{figure}[htb]
\begin{center}
\fbox{
\parbox{6in}{
\begin{tabbing}
AAA\=AAA\=AAA\=AAA\=AAA\=AAA\kill
{\bf Algorithm} \>\>\> {\sf Create\_Content\_Definition}\\
{\bf Input}     \>\>\> An element $e\in \E_{d_R}$ \\
		\>\>\> A query $Q$ with nodes $N_Q$ and projected nodes $O$ \\
		\>\>\> The originating DTD $d$ with content definition 
$\varphi_e$ for $e$ \\
{\bf Output}    \>\>\> The content definition of $e$ in the result DTD\\
\\
{\bf If} $(\exists n_O\in O)$ s.t.\ $($\=$(l(n_O) = e)$ {\bf or} \\
			\>  $(l(n_O) = e'$ {\bf and} $dtd\_desc(e',e)))$\\
AAA\=AAA\=AAA\=AA\=AA\=AAA\kill
	\> {\bf then} $\varphi := \varphi_e$ \\
{\bf Else} $\varphi := \emptyset$ \\
{\bf For all} nodes $n_Q\in N_Q$ {\bf do} \\
	\> {\bf If} $((l(n_Q) = e)$ {\bf and} ($anc(n_Q,O)$))
				{\bf then}\\
	\>  \>$\varphi_{n_Q} := \varphi_e$\\
       \>\> {\bf For all} elements $e'$ in $\varphi_e$ {\bf do}\\
       \>\>\>{\bf If} there is a child of $n_Q$, namely $n'_Q$ s.t.\\
	\>\>\>\> (($l(n'_Q) = e'$) {\bf and} ($anc(n'_Q,O)))$ {\bf then} \\
       \>\>\>\>\>{\bf Replace} all occurrences of $e'$ in $\varphi_{n_Q}$ with $(e'?)$\\
       \>\>\>{\bf Else}
       {\bf Replace} all occurrences of $e'$ in $\varphi_{n_Q}$ with $\emptyset$\\
       \>$\varphi := \varphi \mid \varphi_{n_Q}$\\
{\bf Return} {\sf Simplify($\varphi$)}.
\end{tabbing}
}}
\end{center}
\caption{Content Definition Generation Algorithm}
\label{CDD}
\end{figure}

\begin{theorem}[Correctness of DTD Creation] \label{Theorem-Compute:DTD}
Let $Q$ be a query with an originating DTD $d$ 
and let $X$ be a document. Suppose that the result of evaluating $Q$ on $X$ is
the XML document $R$. Then $R$ strictly conforms to the result DTD as 
formed by the process described above.
In addition, the computation of the result DTD can be performed in time 
$O(|d| + |Q|)$.
\end{theorem}

Note that it follows from Theorem~\ref{Theorem-Compute:DTD} that the result
DTD is linear in the size of the original DTD.  The compactness of
the result DTD makes the requerying process simpler, since requerying entails
exploring the result DTD. 

According to Theorem~\ref{Theorem-Compute:DTD}, the resulting documents
conform to the result DTD. The question arises as to how precisely the 
result DTD describes the resulting documents. In order to answer this question
we define a partial order on DTDs (\cite{PV:DTD}). 
Given a DTD $d$ we denote the 
set of XML documents that strictly conform to $d$ as ${\sl conf}(d)$.
Given DTDs $d$ and $d'$ we say that $d$ is {\em tighter\/} than $d'$,
denoted $d\preceq d'$, if ${\sl conf}(d)\subseteq {\sl conf}(d')$.
We say that $d$ is {\em strictly tighter\/} that $d'$, denoted 
$d\prec d'$, if ${\sl conf}(d)\subset {\sl conf}(d')$.

Intuitively, it would be desirable to find a result DTD $d_R$ that is as tight
 as possible, under the restriction that all possible result documents must
strictly conform to $d_R$. 
However, our algorithm does not necessarily find the tightest possible result 
DTD. In other words, our algorithm may create a result DTD $d_R$ although
there exists a DTD $d'_R$ to which all resulting documents must strictly
conform and $d'_R\prec d_R$. If $d_R$ is the tightest possible result DTD,
we call $d_R$ a {\em minimal result DTD\/}.
According to the following Proposition, there is a query and DTD for which 
a minimal result DTD must be exponential in the size of the original DTD. 

\begin{proposition}[Exponential Result DTD]
There is a query $Q$ created from an originating DTD $d$, such that if $d'$ is
a minimal result DTD of $Q$, then $d'$ is of size $\O(\,|d|!\,)$.
\end{proposition}
	

Observe that an
exponential blowup of the original DTD is undesirable for two reasons.
First, creating such a DTD is intractable. Second, if the result DTD
is of exponential size, then it is difficult for a user to requery
previous results. Thus, our algorithm for creating result DTDs
actually returns a convenient DTD, although it is not always minimal.

\section{Correctness of {\sf Query\_Evaluate}} \label{Section-Proofs}

In this Section we prove the correctness of the algorithm 
{\sf Query\_Evaluate} presented in Figure~\ref{Figure:Evaluation}.
We first prove some necessary lemmas.

\begin{lemma} \label{Lemma-Paths:Equal}
Let $n_X$ be a node in a document $X$ and let $n_Q$ be a node in a query 
$Q$. If there exists a matching $\mu$ of $X$ to $Q$ such that $n_X\in \mu(n_Q)$
then {\sl path$(n_X)$ =  path$(n_Q)$\/}.
\end{lemma}

\proof
Suppose that $n_X\in\mu(n_Q)$. We show by induction on the depth of $n_Q$
that {\sl path$(n_X)$ =  path$(n_Q)$\/}. Suppose that the labeling function 
of $Q$ is $l_Q$ and that the labeling function of $X$ is $l_X$. \\
{\it Case 1:\/} Suppose that $n_Q$ is the root of $Q$. According 
to Condition~\ref{Matching:Label} in Definition~\ref{Definition-Matching},
it holds that
$l_Q(n_Q) = l_X(n_X)$. According to Condition~\ref{Matching:Roots} in 
Definition~\ref{Definition-Matching},
$n_X$ is the root of $X$. Thus, clearly the claim holds.  \\
{\it Case 2:\/} Suppose that $n_Q$ is of depth $m$. Once again, according 
to Condition~\ref{Matching:Label} in Definition~\ref{Definition-Matching}, 
it holds that
$l_Q(n_Q) = l_X(n_X)$. In addition, $p(n_X)\in\mu(p(n_Q))$
(see Condition~\ref{Matching:Connected} in 
Definition~\ref{Definition-Matching}). Note that $p(n_Q)$ is of depth $m-1$.
Thus, by the induction hypothesis, {\sl path$(p(n_X))$ =  path$(p(n_Q))$\/}. 
Our claim follows. 
$\qed$

We present an auxiliary definition. 
We define the {\em height\/} of a query node $n_Q$, denoted $h(n_Q)$, as 
\[           h(n_Q) = \left\{ \begin{array}{ll}
			0 & \mbox{if $n_Q$ is a leaf node}\\
			{\sl max}\set{h(n'_Q)|\mbox{$n'_Q$ is a child of $n_Q$}} + 1 
						& \mbox{otherwise}
			\end{array}
	\right. \]

We show that the algorithm implicitly defines a satisfying
matching $\mu_R$. The nodes that are returned are those corresponding to output 
nodes in $\mu_R$, and their ancestors and descendents. 

We define the function $\mu_R:N_Q\rightarrow 2^{N_X}$ in the following way:
\[ \mbox{$n_X\in\mu_R(n_Q)$ $\iff$ {\sl match\_array}[$n_Q$,$n_X$]= ``true''}\]
Note that we consider the values of {\sl match\_array} at the end of the
evaluation of {\sf Query\_Evaluate}. We call $\mu_R$ the 
{\em retrieval function\/} of {\sf Query\_Evaluate} w.r.t.\ $X$ and $Q$.

\begin{lemma}[Retrieval Function is a Satisfying Matching]\label{Lemma-Satisfying:Matching}
Let $X$ be a document, let $Q$ be a query and let $\mu_R$ be the retrieval 
function of {\sf Query\_Evaluate} w.r.t.\ $X$ and $Q$. Then $\mu_R$ 
is a satisfying matching of $X$ to $Q$.
\end{lemma}

\proof 
We show that $\mu_R$ is a matching, i.e., that $\mu_R$ meets the conditions in 
Definition~\ref{Definition-Matching}.
\begin{itemize}
	\item {\bf Roots Match:} The only node that has the same path as 
$r_Q$ is $r_X$. Thus, the only time that the function {\sf Matches} is called 
for the root of the query is with the root of the document. 
Thus, the value of $\mu_R(r_Q)$ must either be either $\set{r_X}$ or $
\emptyset$. 
However, it is easy to see that if $\mu_R(r_Q) = \emptyset$ then {\sf Query\_Evaluate} returns $\emptyset$. Thus, it must hold that $\mu_R(r_Q) = \set{r_X}$.
	\item {\bf Node Matching:} If $n_X\in \mu_R(n_Q)$ then {\sf Matches} was called with 
$n_Q$ and $n_X$. Thus $n_X$ and $n_Q$ have the same path, and hence, $n_X$
matches $n_Q$.
	\item {\bf Connectivity:} If {\sl match\_array}[$p(n_Q)$,$p(n_X)$] does not hold, then {\sl match\_array}[$n_Q$,$n_X$] is 
assigned the value ``false''. Therefore, clearly the connectivity requirement
of a matching holds.
\end{itemize}

We now show that $\mu_R$ is a satisfying matching (see 
Definition~\ref{Definition-Satisfying:Matching}). Suppose that $n_X\in\mu_R(n_Q)$
for a document node $n_X$ and a query node $n_Q$. Note that this 
implies that {\sf Matches} returned the value ``true'' when applied to $n_Q$
and $n_X$. We prove by induction
on the height of $n_Q$ that the appropriate condition holds. 
We consider three cases.
\begin{itemize}
	\item Suppose that $h(n_Q) = 0$. Then $n_Q$ is a leaf and 
$c(n_Q)(t(n_X))$ must hold as required.
	\item Suppose that $h(n_Q)>0$ and that 
$n_Q$ is an or-node. The procedure {\sf Matches} returned the value ``true'' 
when applied to $n_Q$ and $n_X$. Therefore, one of the following must hold:
	\begin{enumerate}
		\item  The condition $c(n_Q)(t(n_X))= \top$ holds.
		\item  At the time of application of {\sf Matches}
to $n_Q$ and $n_X$, there was a child $m_Q$ of $n_Q$ and a child $m_X$
of $n_X$ such that {\sl match\_array}[$m_Q$,$m_X$] had the value ``true''.
Note that it follows that $m_X$ matches $m_Q$.  
The value of  {\sl match\_array}[$m_Q$,$m_X$] will not be changed to ``false''
during the evaluation of {\sf Query\_Evaluate} since 
{\sl match\_array}[$n_Q$,$n_X$] is ``true''. Thus, $m_X\in\mu_R(m_Q)$.
	\end{enumerate}
In either case Condition~\ref{def:satisfying:matching:or}  from Definition~\ref{Definition-Satisfying:Matching} holds as required.
	\item  Suppose that $h(n_Q)>0$ and that 
$n_Q$ is an and-node. We omit the proof as it is similar to the previous case.
\end{itemize}
Thus, $\mu_R$ is a satisfying matching as required.
\qed

We say that a matching $\mu$ {\em contains\/} a matching $m'$ if there exists
a matching $\mu''$ such that $\mu = \mu' \cup \mu''$. We will show that
the retrieval function contains all other satisfying matchings.

\begin{lemma}[Retrieval Function Containment]\label{Lemma-Containment}
Let $X$ be a document, let $Q$ be a query and let $\mu_R$ be the retrieval 
function of {\sf Query\_Evaluate} w.r.t.\ $X$ and $Q$. Suppose that $\mu$
is a satisfying matching of $X$ to $Q$. Then $\mu_R$ contains $\mu$. 
\end{lemma}

\proof
Suppose that $\mu$ is a satisfying matching. We show by induction on the 
height of $n_Q$ that $\mu(n_Q)\subseteq \mu_R(n_Q)$. We first consider the 
values assigned to {\sl match\_array} during the first pass (the bottom-up pass) of the algorithm. 
\begin{itemize}
	\item Suppose that $h(n_Q) = 0$. Suppose that $n_X\in \mu_R(n_Q)$. Then,
according to Lemma~\ref{Lemma-Paths:Equal} {\sl path}($n_X$) = {\sl path}($n_Q$). Thus, {\sf Matches} will be called on $n_X$ and $n_Q$. The condition 
$c(n_Q)(t(n_X))$ holds. Thus, {\sl match\_array}[$n_Q$,$n_X$] will be assigned
the value ``true''.
	\item Suppose that $h(n_Q) > 0$ and $n_Q$ is an or-node. Suppose that $n_X\in \mu_R(n_Q)$. Once again, 
according to Lemma~\ref{Lemma-Paths:Equal} {\sl path}($n_X$) = {\sl path}($n_Q$). Thus, {\sf Matches} will be called on $n_X$ and $n_Q$. It also follows 
that one of the following must hold:
	\begin{enumerate}
		\item The value of $c(n_Q)(t(n_X))$  is ``true''. Thus, 
{\sf Matches} returns true.
		\item There is a child $m_Q$ of $n_Q$ and a child $m_X$ of 
$n_X$ such that $m_X$ matches $m_Q$ and $m_X\in\mu(m_Q)$. Note that 
$h(m_Q)<h(n_Q)$. Thus, by the induction hypothesis, the value of {\sl match\_array}[$m_Q$,$m_X$] after the first pass of the algorithm is ``true''.  Thus
{\sf Matches} returns true when called on $n_Q$ and $n_X$.
	\end{enumerate}
	\item Suppose that $h(n_Q) > 0$ and $n_Q$ is an and-node. We omit
the proof as it is similar to the previous case. 
\end{itemize}
It is easy to see that it follows from the connectivity of $\mu$ that if $n_X\in\mu(n_Q)$ then
 the value of {\sl match\_array}[$n_Q$,$n_X$] will not be changed to ``false''.
Thus, $\mu$ is contained in $\mu_R$ as required.
\qed

We can now prove the theorem required.

\begin{theorem}[Correctness of {\sf Query\_Evaluate}]
Given document $X$ and a query $Q$, the algorithm {\sf Query\_Evaluate}
retrieves the output set of $X$ w.r.t.\ $Q$.
\end{theorem}

\proof 
Let $X$ be a document and $Q$ be a query. We show that a
document node $n_X$ is returned by {\sf Query\_Evaluate} if and only
if $n_X$ is in the output set of $X$ w.r.t.\ $Q$.\\

``$\Leftarrow$'' Suppose that $n_X$ is in the output set of $X$ w.r.t.\
$Q$. Then there is a satisfying matching $\mu$ of $X$ to $Q$ such that 
and an output node $n_Q$ in $Q$ such that either $n_X\in\mu(n_Q)$
or $n_X$ is an ancestor or descendent of a node in $\mu(n_Q)$. Let $\mu_R$
be the retrieval function of {\sf Query\_Evaluate} w.r.t.\ $X$ and $Q$. 
According to Lemma~\ref{Lemma-Containment} $\mu$ is contained in $\mu_R$. 
Thus, clearly $n_X$ is returned by {\sf Query\_Evaluate}.

``$\Rightarrow$'' Suppose that $n_X$ is returned by {\sf Query\_Evaluate}. 
Let $\mu_R$
be the retrieval function of {\sf Query\_Evaluate} w.r.t.\ $X$ and $Q$.
Then there is an output node $n_Q$ in $Q$ such that either $n_X\in\mu_R(n_Q)$
or $n_X$ is an ancestor or descendent of a node in $\mu_R(n_Q)$. According
to Lemma~\ref{Lemma-Satisfying:Matching} $\mu_R$ is a satisfying matching
of $X$ to $Q$. Thus, $n_X$ is in the output set of $X$ w.r.t.\ $Q$.
\qed

\nocite{Abiteboul:Et:Al-The:Lorel:Language-IJDL}
\nocite{xml-lorel:Web-DB}
\nocite{XML:Book}
\nocite{Goldman:Widom-Interactive:Query-WebDB}
\nocite{XML-Namespaces}
\nocite{XML-Schema}
\nocite{Bar-Yossef:Et:Al-Querying:Semantically:Tagged:Documents:on:the:Web-NGITS99}
\nocite{Kanza:Nutt:Sagiv-Queries:with:Incomplete:Answers:over:SSD-PODS99}

\bibliography{%
        main}
\bibliographystyle{alpha}

\end{document}